\documentclass[useAMS,usenatbib]{mn2e}
\usepackage{amssymb,amsmath}
\usepackage{graphicx}
\usepackage{epstopdf}

\def\UW{$^{1}$}
\def\Curtin{$^{2}$}
\def\USwinburne{$^{3}$}
\def\CfA{$^{4}$}
\def\ASU{$^{5}$}
\def\ANU{$^{6}$}
\def\CAASTRO{$^{7}$}
\def\CSIRO{$^{8}$}
\def\Haystack{$^{9}$}
\def\RRI{$^{10}$}
\def\USydney{$^{11}$}
\def\MIT{$^{12}$}
\def\VUW{$^{13}$}
\def\UWisc{$^{14}$}
\def\UMelbourne{$^{15}$}
\def\UTasmania{$^{16}$}
\def\PerthUWA{$^{17}$}

\begin{document}

\title[MWA Sensitivity]{The EoR Sensitivity of the Murchison Widefield Array}

\author[A.~P.~Beardsley, et al.]{
A.~P.~Beardsley\UW,  
B.~J.~Hazelton\UW, 
M.~F.~Morales\UW\thanks{Email: mmorales@phys.washington.edu},
W.~Arcus\Curtin, 
D.~Barnes\USwinburne,
\newauthor  
G.~Bernardi\CfA, 
J.~D.~Bowman\ASU, 
F.~H.~Briggs\ANU$^,$\CAASTRO, 
J.~D.~Bunton\CSIRO, 
R.~J.~Cappallo\Haystack, 
\newauthor
B.~E.~Corey\Haystack, 
A.~Deshpande\RRI, 
L.~deSouza\CSIRO$^,$\USydney, 
D.~Emrich\Curtin, 
B.~M.~Gaensler\USydney$^,$\CAASTRO,  
\newauthor
R.~Goeke\MIT, 
L.~J.~Greenhill\CfA, 
D.~Herne\Curtin, 
J.~N.~Hewitt\MIT, 
M.~Johnston-Hollitt\VUW, 
\newauthor
D.~L.~Kaplan\UWisc, 
J.~C.~Kasper\CfA, 
B.~B.~Kincaid\Haystack,  
R.~Koenig\CSIRO, 
E.~Kratzenberg\Haystack, 
\newauthor
C.~J.~Lonsdale\Haystack,  
M.~J.~Lynch\Curtin, 
S.~R.~McWhirter\Haystack, 
D.~A.~Mitchell\UMelbourne$^,$\CAASTRO,  
\newauthor
E.~Morgan\MIT, 
D.~Oberoi\Haystack,  
S.~M.~Ord\CfA, 
J.~Pathikulangara\CSIRO,  
T.~Prabu\RRI, 
\newauthor
R.~A.~Remillard\MIT, 
A.~E.~E.~Rogers\Haystack,  
A.~Roshi\RRI, 
J.~E.~Salah\Haystack, 
R.~J.~Sault\UMelbourne, 
\newauthor
N.~Udaya~Shankar\RRI, 
K.~S.~Srivani\RRI, 
J.~Stevens\CSIRO$^,$\UTasmania,  
R.~Subrahmanyan\RRI$^,$\CAASTRO, 
\newauthor
S.~J.~Tingay\Curtin$^,$\CAASTRO,  
R.~B.~Wayth\Curtin$^,$\CfA$^,$\CAASTRO, 
M.~Waterson\Curtin$^,$\ANU, 
R.~L.~Webster\CAASTRO$^,$\UMelbourne, 
\newauthor
A.~R.~Whitney\Haystack, 
A.~Williams\PerthUWA, 
C.~L.~Williams\MIT, 
and
J.~S.~B.~Wyithe\CAASTRO$^,$\UMelbourne
\\
$^{1}$University of Washington, Seattle, USA\\
$^{2}$International Centre for Radio Astronomy Research, Curtin University, Perth, WA 6845, Australia\\
$^{3}$Swinburne University of Technology, Melbourne, Australia\\
$^{4}$Harvard-Smithsonian Center for Astrophysics, Cambridge, USA\\
$^{5}$Arizona State University\\
$^{6}$The Australian National University, Canberra, Australia\\
$^{7}$ARC Centre of Excellence for All-sky Astrophysics (CAASTRO)\\
$^{8}$CSIRO Astronomy and Space Science, Australia\\
$^{9}$MIT Haystack Observatory, Westford, USA\\
$^{10}$Raman Research Institute, Bangalore, India\\
$^{11}$University of Sydney, Sydney, Australia\\
$^{12}$MIT Kavli Institute for Astrophysics and Space Research, Cambridge, USA\\
$^{13}$Victoria University of Wellington, New Zealand\\
$^{14}$University of Wisconsin--Milwaukee, Milwaukee, USA\\
$^{15}$The University of Melbourne, Melbourne, Australia\\
$^{16}$University of Tasmania, Hobart, Australia\\
$^{17}$Perth Observatory, Perth, Australia, and the University of Western Australia\\
}

\label{firstpage}
\pagerange{\pageref{firstpage}--\pageref{lastpage}}\pubyear{2012}

\maketitle

\begin{abstract}
Using the final 128 antenna locations of the Murchison Widefield Array (MWA), we calculate its sensitivity to the Epoch of Reionization (EoR) power spectrum of redshifted 21 cm emission for a fiducial model and provide the tools to calculate the sensitivity for any model.  Our calculation takes into account synthesis rotation, chromatic and asymmetrical baseline effects, and excludes modes that will be contaminated by foreground subtraction.  For the fiducial model, the MWA will be capable of a 14$\sigma$ detection of the EoR signal with one full season of observation on two fields (900 and 700 hours).
\end{abstract}

\clearpage

\section{Introduction}
The Cosmic Dark Ages and the Epoch of Reionization (EoR) remain a largely unexamined chapter of the history and evolution of the Universe.  Observation of redshifted 21 cm emission shows promise of probing the EoR (see \citealt{Furlanetto:2006} and \citealt{Morales:2010} for recent reviews).  Indeed, studies of the EoR and dark energy were rated with high priority by the 2010 Astronomy and Astrophysics Decadal Survey.  Several ground-based radio experiments are currently under construction to probe the EoR through 21 cm power spectrum measurements, including LOFAR (LOw Frequency Array\footnote{http://www.lofar.org/}), PAPER (Precision Array for Probing the Epoch of Reionization\footnote{http://eor.berkeley.edu/}), and the MWA (Murchison Widefield Array\footnote{http://www.mwatelescope.org/}).  

The MWA is being built in the radio quiet Murchison Radio Observatory in Western Australia, and aims to measure the EoR power spectrum via the 21 cm signal over a large range of redshifts.  The originally planned MWA was to consist of 512 antennas, distributed over a circular region of radius 1.5 km \citep{Lonsdale:2009}. With current funding the instrument has been re-scoped to 128 antennas, but will have similar layout characteristics to the originally planned 512 antenna array. A full description of the 128 antenna instrument is presented in \citet{Tingay:2012}, and a thorough description of the science capabilities will be presented in \citet{Bowman:2012}.

\begin{figure}
\begin{center}
\includegraphics[width=\columnwidth]{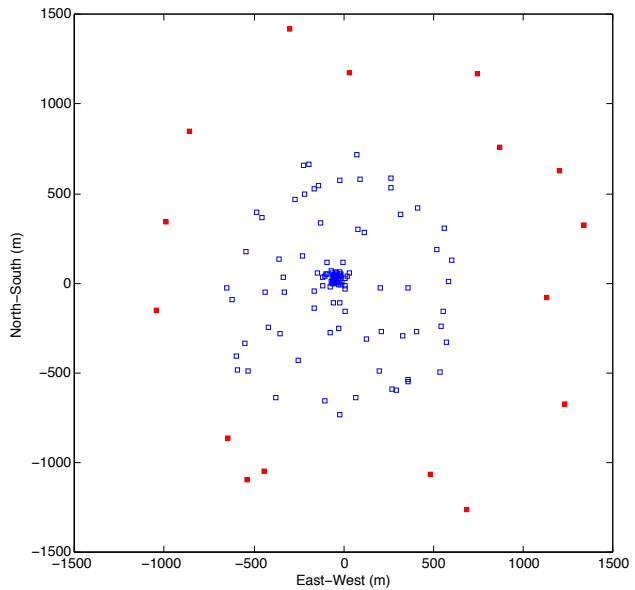}\label{fig:layout}
\caption{The antenna locations for the 128 antenna MWA.  Positions are measured relative to -26$^{\circ}$ 42' 4.396" Latitude, 116$^{\circ}$ 40' 13.646" Longitude. The blue squares show the core 112 antennas which will be integrated for an EoR measurement. The solid red squares represent the outlier antennas used for solar measurements, but are not used for EoR measurements.  While the antennas are indeed square, the squares shown here are not to scale.}
\end{center}
\end{figure}

Here we calculate the MWA's expected sensitivity to the EoR signal using the physical antenna locations. A given baseline (the separation vector between any two antennas) is sensitive to a particular angular Fourier mode on the sky, so the baseline distribution is directly related to the EoR sensitivity of an array \citep{Morales:2005}. The MWA baseline distribution will have a dense core for EoR sensitivity and a smooth extended radial profile for calibration and foreground subtraction purposes \citep{Bowman:2006}. The locations of the 128 antennas for the MWA were optimized using the algorithm presented in \citet{Beardsley:2012} and shown in Fig. \ref{fig:layout}. A table of the locations of all 128 antennas are available in the electronic supplement.

We use a fiducial model to calculate the MWA's sensitivity and in attached tables provide the information needed to quickly apply any model. The EoR observing plan for the MWA is to track fields when they are above 45 degrees elevation and the sun and galactic center are below the horizon.  Over an annual cycle, this yields a full observational season of 900 hours integration on a primary field and 700 hours on a second field.  For the fiducial model, we find that with a full season of observation the MWA will be capable of a 14$\sigma$ power spectrum detection, along with constraints on the slope.

Throughout this paper we use a $\Lambda$CDM cosmology with $\Omega_m = 0.73$, $\Omega_{\Lambda} = 0.27$, and $h = 0.7$, consistent with WMAP seven year results \citep{Komatsu:2011}.  All distances and wavenumbers are in comoving coordinates.

\section{EoR Sensitivity}\label{sec:sensitivity}
The power spectrum measurement of the sky temperature is done in three dimensions (two angular directions, and the line-of-sight direction achieved through redshift), so we must find the uncertainty in each three dimensional voxel in cosmological wavenumber ($\mathbf{k}$) space, then perform a weighted average in spherical bins to arrive at one-dimensional sensitivity (following \citealt{Morales:2005}, \citealt{McQuinn:2006}, and \citealt{Morales:2010}).

The fundamental visibility measurement of an interferometer is done in ($u,v,f$) space, where $u$ and $v$ are the baseline coordinates (measured in wavelengths), and $f$ is the frequency of the observation.  The thermal uncertainty on the visibility measurement is given by  
\begin{equation}\label{eq:thermal_var}
V_{\text{rms}}(u,v,f)  = \frac{c^2 T_{\text{sys}}}{f^2A_e \sqrt{\Delta f \tau}},
\end{equation}
where $T_{\text{sys}}$ is the system temperature, $A_e$ is the effective collecting area per antenna, $\Delta f$ is the frequency channel width, and $\tau$ is the total integration time for the mode including redundant baselines \citep{Morales:2010}.  Observational parameters for our calculation are shown in Table \ref{tbl:obs_params}.  The system temperature is dominated by galactic foreground emission, and redshift dependence is discussed in \citet{Bowman:2006}.  Here we assume a constant system temperature over the observational bandwidth.

%\begin{table}
%\begin{center}
%\caption{Observational parameters for sensitivity estimation\label{tbl:obs_params}}
%\begin{tabular}{lc}
%\hline
%Parameters & Values \\
%\hline
%No. of antennas & 112*\\
%Central frequency & 158 MHz (z $\sim$ 8)\\
%Field of view & 31$^{\circ}$ \\
%Effective area per antenna & 14.5 m$^2$\\
%Total bandwidth & 8 MHz\\
%T$_{\mathrm{sys}}$ & 440 K\\
%Channel width & 40 kHz\\
%Latitude & -26.701$^{\circ}$\\
%Primary Field RA & 6$^\text{h}$\\
%Secondary Field RA & 0$^\text{h}$\\
%\hline
%\multicolumn{2}{l}{*Sixteen of the 128 antennas are not integrated for EoR}\\
%\multicolumn{2}{l}{measurements and are not included here.}
%\end{tabular}
%\end{center}
%\end{table}

\begin{table}
\begin{center}
\caption{Observational parameters for sensitivity estimation\label{tbl:obs_params}}
\begin{tabular}{lc}
\hline
Parameters & Values \\
\hline
No. of antennas & 112*\\
Central frequency & 158 MHz (z $\sim$ 8)\\
Field of view & 31$^{\circ}$ \\
Effective area per antenna & 14.5 m$^2$\\
Total bandwidth & 8 MHz\\
T$_{\mathrm{sys}}$ & 440 K\\
Channel width & 40 kHz\\
Latitude & -26.701$^{\circ}$\\
Primary Field RA & 6$^\text{h}$\\
Secondary Field RA & 0$^\text{h}$\\
\hline
\multicolumn{2}{p{6.5cm}}{*Sixteen of the 128 antennas are not integrated for EoR measurements and are not included here.}
\end{tabular}
\end{center}
\end{table}

%\begin{deluxetable}{lc}
%%\tabletypesize{\scriptsize}
%\tablecaption{Observational parameters for sensitivity estimation\label{tbl:obs_params}}
%\tablewidth{0pt}
%\tablehead{\colhead{Parameters} & Values}
%\startdata
%No. of antennas & 112\tablenotemark{*}\\
%Central frequency & 158 MHz (z $\sim$ 8)\\
%Field of view & 31$^{\circ}$ \\
%Effective area per antenna & 14.5 m$^2$\\
%Total bandwidth & 8 MHz\\
%T$_{\mathrm{sys}}$ & 440 K\\
%Channel width & 40 kHz\\
%Latitude & -26.701$^{\circ}$\\
%Primary Field RA & 6$^\text{h}$\\
%Secondary Field RA & 0$^\text{h}$
%\enddata
%\tablenotetext{*}{Sixteen of the 128 antennas are not integrated for EoR measurements and are not included here.} 
%\end{deluxetable}

%The effective area estimate used here corresponds to an ideal dipole approximation.  Simulations and preliminary measurements of the actual MWA antenna tiles indicate that the zenith gain will be $\sim$20 m$^2$, but since the EoR fields will be observed between zenith angles of 0 and 45 degrees, we used the conservative value, which roughly corresponds to the gain at 45 degrees for the actual tiles.

To determine $\tau$, the integration time per $(u,v,f)$ voxel, we use the surveyed antenna locations, and perform an aperture rotation for 3 hours on either side of zenith.  We approximate chromatic effects by calculating the baseline migration along the frequency dimension, then averaging.  This includes chromatic effects while avoiding a full covariance calculation (Hazelton, et al. 2012, in preparation).  

\begin{figure}
\begin{center}
\includegraphics[width=\columnwidth]{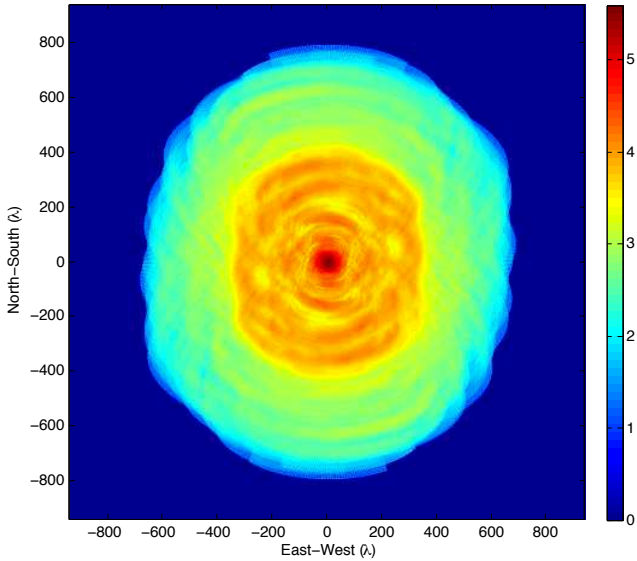}
\caption{Effective integration time per day per ($u,v$) cell including rotation synthesis for 128 antenna MWA at $z \sim 8$ ($\lambda = 1.89$ m).  The color scale units are the logarithm of effective seconds observed per day per angular mode, assuming six hours of integration per day on one EoR field.  Note that the total number of seconds observed per day is 21600, but an angular mode can be effectively observed longer due to redundant baselines.  The most observed mode in this array is $\sim4 \times 10^5$ sec/day.  The \emph{uv} cell size is dictated by the size of the instrumental window function.  
Following \citet{Bowman:2006} we used a cell size of $(8.3$ m$)^2$. Data for this figure will be available in a machine readable table to easily plug into Equation \ref{eq:thermal_var}.}
\label{fig:uv}
\end{center}
\end{figure}

The sampling matrix for one day of observation on a single EoR field is shown in Fig. \ref{fig:uv}.    The MWA will have a very dense, highly redundant $uv$ core, with a smooth radial profile extending to 1.5 km.  The large number of baselines in the core will beat down the thermal variance for those modes because the effective observing time is the sum of all the baselines observing the mode.

The $(u,v)$ coordinates map directly to the transverse cosmological wavenumber by the relation $k_{\perp}=2\pi u/D$, where $D$ is the comoving distance to the observation.  The observing frequency dimension maps to the line-of-sight direction, and must be Fourier transformed to the $k_{||}$ dimension.  Once these conversions are done, our data is in three dimensional $k$-space, and we square to reach the power spectrum.  Propagating errors, the thermal uncertainty per k-space bin is given by
\begin{equation}\label{eq:thermal_var}
C^N(\mathbf{k}) = T_{\text{sys}}^2\left(\frac{D^2\lambda^2}{A_e}\right)\left(\frac{\Delta D}{B}\right)\frac{1}{\tau}.
%C^N(\mathbf{k}) = \left(\frac{D^2\lambda^2}{A_e}\right)\left(\frac{\Delta D}{B}\right)\left(\frac{T_{\text{sys}}^2}{\tau}\right).
%\frac{T_{\text{sys}}^2 D^2 \Delta D \lambda^2}{B A_e} \frac{1}{\tau}
\end{equation}
The second term can be thought of as converting the $uv$ bin size ($A_e$) to cosmological wavenumber space and has units of Mpc$^2$. The third term converts the width of the observation from bandwidth to line of sight spatial extent and has units of Mpc s (for flat space the line-of-sight and transverse distances are equivalent), and $\tau$ is the integration time for the k-space bin (in seconds).  Inserting the values from Table \ref{tbl:obs_params} for all terms except the integration time gives 
\begin{equation}\label{eq:thermal_var}
C^N(\mathbf{k}) = \frac{6.95 \times 10^7}{\tau}\ \rm{mK}^2 \rm{Mpc}^3.
\end{equation}

There is also a sample variance contribution to the uncertainty.  Assuming the distribution is Gaussian, the sample variance per three dimensional voxel is given by the power spectrum itself \citep{McQuinn:2006}.  Combining the thermal and sample uncertainties gives  the total variance per 3D $k$-space voxel
\begin{equation}\label{eq:total_var}
\sigma_P^2(\mathbf{k}) = \left(P_{21}(\mathbf{k}) + C^N(\mathbf{k})\right)^2.
\end{equation}

Because of the sample variance term, the calculated sensitivity of an array depends on one's choice of theoretical EoR model. While surveying the landscape of EoR models is beyond the scope of this paper, we have included a table of the effective seconds observed per day per ($u,v$) cell in the electronic supplement (data for Figure \ref{fig:uv}). The seconds per day can be combined with the observing strategy to calculate the integration time per cell, $\tau$ in Equation \ref{eq:thermal_var}, and combined with the theoretical model in Equation \ref{eq:total_var} to accurately determine the  sensitivity of the MWA for any proposed model. The coefficient values in Equation \ref{eq:thermal_var} and the coordinates of the supplemental table can be scaled to different redshifts with $\sim5\%$ error on the resulting sensitivity, or the antenna locations and synthesis rotation can be used to recalculate the integration time per bin using the supplemental table as a  cross-check. In the remainder of this paper we use the fiducial model of a fully neutral IGM \citep{Furlanetto:2006}\footnote{Available online at www.astro.ucla.edu/~sfurlane/21cm\_pk.htm} as an example of how to accurately calculate the EoR power spectrum sensitivity. 

The underlying EoR fluctuations are assumed to be isotropic, however velocity distortions will amplify the signal in the line-of-sight direction on relevant large scales \citep{Barkana:2005}. For our fiducial model this angular dependence is  given by $P_{21}(\mathbf{k}) = (1+2\mu^2+\mu^4)P_{21}(k)$, where $\mu = k_{||}/|\mathbf{k}|$.  This effect depends on whether dark matter or ionizing sources are sourcing the fluctuations. Throughout reionization both sources will be relevant and the above expression will depend on the cross-power spectrum between the fluctuations. Because our fiducial model is a fully neutral IGM we can use this simplified relation.  In addition the MWA is sensitive to much smaller $k_{\perp}$ modes compared to $k_{||}$ modes,  so this effect is significant for the dark matter sourced fiducial model.

\begin{figure}
\begin{center}
\includegraphics[width=\columnwidth]{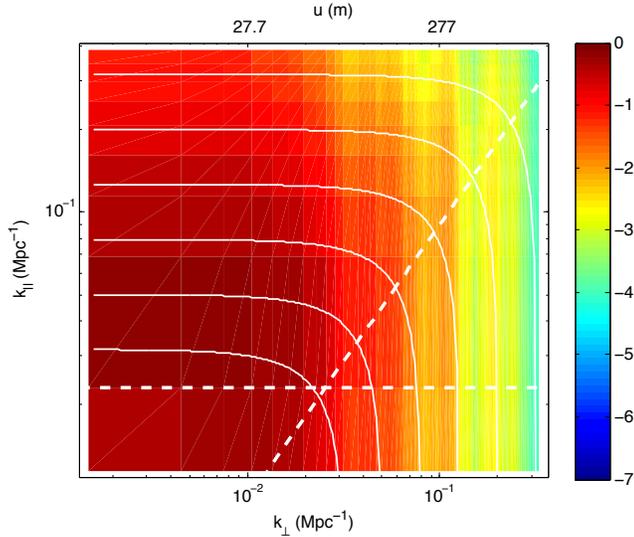}
\caption{Estimated power spectrum sensitivity to EoR signal per voxel for the MWA.  The quantity plotted is $\log_{10}(P_{21}(\mathbf{k})/(\sigma_P(\mathbf{k}))$ for a two dimensional slice of the three dimensional data cube with 900 hours of integration.  The white curved lines show the bin edges used for the one dimensional plot (Fig. \ref{fig:1D}).  The data below the horizontal dashed line and to the right of the diagonal dashed line will be contaminated by foregrounds. Only data within the EoR window (the upper left) is used to calculate the sensitivity in Fig. \ref{fig:1D}. For reference, the corresponding baseline lengths are given on the top axis.}
\label{fig:2D}
\end{center}
\end{figure}

Figure \ref{fig:2D} shows the signal to noise per voxel in a slice through the 3D $k$-space.  At low $k$, a large signal and dense baseline distribution result in a signal to noise approaching 1.  Moving up in $k_{||}$, the signal diminishes, but the baseline density remains constant, so the sensitivity drops relatively slowly.  Moving up in $k_{\perp}$, however, both the signal and the baseline density drop, resulting in a more drastic drop in sensitivity.

Foreground contamination limits the observability of the EoR.  Fortunately, the contamination is localized in 3D $k$-space, leaving a relatively uncontaminated EoR window \citep{Vedantham:2012, Morales:2012}.  The spectrally smooth foregrounds are fit to low order polynomials over the full 30.72 MHz instrument bandwidth, contaminating low line-of-sight wavenumbers \citep{Bowman:2009}. However, an individual observation is limited to $\sim 8$ MHz due to cosmic evolution, so only our $k_{||}=0$ bin will be contaminated. This exclusion zone is shown in Fig. \ref{fig:2D} by the region below the horizontal white line.  In addition, mode mixing effects will throw power higher in $k_{||}$, creating a wedge shape of contamination \citep{Datta:2011, Morales:2012, Trott:2012}. The location of this contamination is indicated in Fig. \ref{fig:2D} by the region below the diagonal line.  The `EoR window' is to the left of the diagonal line and above the horizontal line.  In this calculation we only use modes within the EoR window.

The next step is to perform a weighted average to condense the three dimensional data into a one dimensional power spectrum.  The underlying power spectrum is expected to be isotropic, so averaging in spherical shells of constant $|\mathbf{k}|$ is appropriate. As discussed earlier, the velocity distortion terms cause the power spectrum to be anisotropic, but can be remedied by dividing the signal and noise by the angular dependence, $(1+2\mu^2+\mu^4)$ in our case.  Then voxels within a constant $k$ shell have the same power spectrum signal, and can be averaged weighting by the uncertainty per voxel.

\begin{figure}
\begin{center}
\includegraphics[width=\columnwidth]{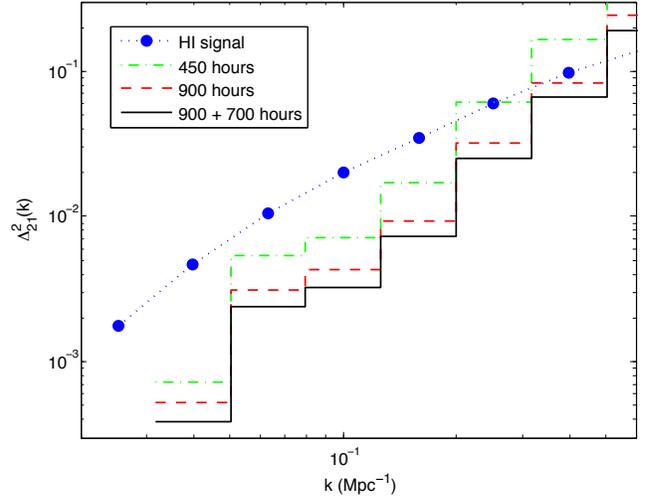}
\caption{Estimated 1D sensitivity for the MWA, for various integration scenarios.  The dotted blue line is the theoretical spherically averaged power spectrum \citep{Furlanetto:2006}, where $\Delta^2_{21}(k) = P_{21}(k)k^3/(2\pi^2T_0^2)$ and $T_0 = 28 [(1+z)/10]^{1/2}$ mK $\approx 26.6$ mK.  The several step functions represent the uncertainty per bin, with the edges of the steps corresponding to the edges of the bins when averaging (white curves in Fig. \ref{fig:2D}). Single field observations are shown for 450 hours (dash-dot green) and 900 hours (dashed red) of integration.  The solid black line corresponds to 900 hours on a primary EoR field, combined with 700 hours on a secondary field.  This averaging excluded any data that would be contaminated by foreground subtraction (below the horizontal line, and to the right of the wedge in Figure \ref{fig:2D}). }
\label{fig:1D}
\end{center}
\end{figure}

Figure \ref{fig:1D} shows the sensitivity of the MWA to this EoR power spectrum.  The theoretical one dimensional spherically averaged power spectrum (dotted blue line) and the uncertainty per bin (various step lines) are plotted.  The uncertainty is plotted as a step function to show the binning used in the spherical average with the edges of the steps corresponding to the white curved lines in Figure \ref{fig:2D}.  

The uncertainty is shown for 450 and 900 hours on one field, as well as a two field observation with 900 hours on one field (RA = 6$^{\text{h}}$) and 700 hours on a second (RA = 0$^{\text{h}}$), corresponding to one full season of observation.  The lowest $k$ bin approaches the sample variance limit as the signal to noise per voxel reaches $\sim 1$ and the array begins to image the largest EoR scales.  The higher $k$ bins, however, are thermal noise dominated at 900 hours.  

We can also follow \cite{Lidz:2008} and fit an amplitude and slope to $\ln\Delta^2_{21}(k)$ in $\ln(k)$,
\begin{equation}
\ln\Delta^2_{21}(k) = \ln\Delta^2_{21}(k=k_p) + \alpha\ln(k/k_p),
\end{equation}
where $k_p$ is a fixed pivot wavenumber.  The uncertainty on the amplitude depends on the pivot wavenumber, and we choose $k_p=0.06$ Mpc$^{-1}$.  The uncertainly is estimated assuming Gaussian statistics, and we fit directly in the 3D $k$-space to avoid binning effects and biases.   For a full season of observation (900 hours on a primary field, 700 hours on a secondary), we predict a  SNR of 14 on the amplitude and 10.9 on the slope ($\alpha$) for the fiducial model. This does not take into account instrument downtime due to inevitable maintenance, nor loss of data for unforeseen reasons.  With a more conservative observation time of 450 hours on a single field, we expect a SNR of 7.1 on the amplitude and 5.0 on the slope.  Even with less than half a full observing season, the MWA has the potential for an EoR detection.

This calculation does not account for systematic biases from calibration and foreground subtraction errors.  Efforts are underway to understand these affects and to achieve this level of sensitivity \citep{Trott:2012}.

\section{Conclusions}
Using the proposed 128 antenna MWA, we have estimated the instrument sensitivity to a model EoR power spectrum, taking into account synthesis rotation, chromatic and asymmetrical baseline effects, and excluding modes that are contaminated by foreground subtraction.  We provide the tools required to calculate the MWA sensitivity for any model.  With an optimistic full season of observation, we would expect to detect the fiducial power spectrum amplitude with SNR $\sim$ 14, and constrain the slope with SNR $\sim 10.9$.  As of mid-July, construction well underway on the MWA, and first light is expected at the end of 2012.

\section*{Acknowledgments}
We acknowledge the Wajarri Yamatji people as the traditional owners of the Observatory site. 

Support came from the U.S. National Science Foundation (grants AST CAREER-0847753, AST-0457585, AST-0908884 and PHY-0835713), the Australian Research Council (LIEF grants LE0775621 and LE0882938), the U.S. Air Force Office of Scientific Research (grant FA9550-0510247), the Centre for All-sky Astrophysics (an Australian Research Council Centre of Excellence funded by grant CE11E0001020), the Smithsonian Astrophysical Observatory, the MIT School of Science, the Raman Research Institute, the Australian National University, the Australian Federal government via the National Collaborative Research Infrastructure Strategy and Astronomy Australia Limited, under contract to Curtin University of Technology, the iVEC Petabyte Data Store, the Initiative in Innovative Computing and NVIDIA sponsored CUDA Center for Excellence at Harvard, and the International Centre for Radio Astronomy Research, a Joint Venture of Curtin University of Technology and The University of Western Australia, funded by the Western Australian State government.

\section*{Supplementary Material}
Two supplementary tables are available for electronic download.  First, \texttt{MWA\_Antenna\_Locations.txt} provides the absolute coordinates of the antennas as they have been placed.  The first column is the antenna number. The second and third columns are the Easting and Northing coordinates for a UTM projection in meters.  The fourth column is the elevation in meters.  All coordinates refer to the south west corner of the antennas.
The second table, \texttt{obs\_time\_table.txt}, provides the expected observed time per \emph{uv} cell per day. This table is the data used to produce Figure \ref{fig:uv}.  The first two columns are the $(u,v)$ coordinates in wavelengths at 158 MHz.  The third column is the observation time per day for the corresponding $uv$ cell in seconds. The grid size is 8.3 m due to the instrumental window function \citep{Bowman:2006}.  The data has been padded with zeros so the array can be reshaped to create a two dimensional image on a regularly spaced grid (see header text). 

\nocite{*}
\bibliography{beardsley}
\bibliographystyle{apj}

\label{lastpage}

\end{document}